\documentclass[amstex,aps,showpacs,preprint]{revtex4}
\usepackage{graphicx}
\usepackage{color}
\usepackage{bm}
\usepackage{longtable}
\usepackage{amsmath}
\usepackage{amsfonts}
\usepackage{amssymb}

\begin{document}

\title{Observing trajectories with weak measurements in quantum systems in
the semiclassical regime}
\author{A.\ Matzkin}
\affiliation{LPTM, CNRS Unit\'{e} 8089, Universit\'{e} de Cergy-Pontoise, 95302
Cergy-Pontoise cedex, France}

\begin{abstract}
We propose a scheme allowing to observe the evolution of a quantum
system in the semiclassical regime along the paths generated by the
propagator. The scheme relies on performing consecutive weak
measurements of the position. We show how ``weak trajectories'' can
be extracted from the pointers of a series of devices having weakly
interacted with the system. The properties of these ``weak
trajectories'' are investigated and illustrated in the case of a
time-dependent model system.
\end{abstract}

\maketitle

In classical physics, the evolution of a physical system is given in terms
of trajectories. Instead quantum mechanics forbids a fundamental description
based on trajectories. Nevertheless the Feynman path integral approach gives
a sum over paths formulation of the evolution of a quantum system, and when
the actions are large relative to $\hslash$ -- the semiclassical regime--,
the wavefunction evolves essentially along classical paths, those of the
corresponding classical system \cite{sc}. Of course, this does not mean that
a quantum object is a localized particle moving on a definite path. But
trajectories may remain significant in quantum systems: the large scale
properties, experimentally observed in many systems \cite{sc,classquant},
display the \emph{signatures} of the underlying classical dynamics.

In this work, we aim to go further by proposing a scheme allowing to observe
the evolution of a quantum system in the semiclassical regime along the
trajectories of the corresponding classical system. The scheme relies on
performing consecutive weak measurements (WM).\ WM \cite{aav,aav2} are
characterized by a very weak coupling between the system and the measurement
apparatus. Thus measuring weakly an observable $\hat{A}$ results in leaving
the former essentially unperturbed while the latter picks up on average a
limited amount of information encapsulated in the weak value (WV)%
\begin{equation}
\left\langle \hat{A}\right\rangle _{W}=\frac{\left\langle \chi \right\vert
\hat{A}\left\vert \psi \right\rangle }{\left\langle \chi \right\vert \left.
\psi \right\rangle };  \label{2}
\end{equation}%
$\left\vert \psi \right\rangle $ is the initial (`preselected') state and $%
\left\langle \chi \right\vert $ is the final (`postselected') state
obtained by performing a standard strong measurement after having
measured $\hat{A}$ weakly. WM are receiving increased attention,
either as a technique for signal amplification \cite{wma} or as a
tool to investigate fundamental problems, from a theoretical
standpoint but also experimentally \cite{wm}. In particular, in a
beautiful recent experiment \cite{at} nonclassical `average
trajectories' for photons deduced indirectly from the WM of momentum
have been observed. In our scheme we introduce instead `weak
trajectories' (WT) by measuring \emph{directly} the position of a
quantum system interacting weakly with a set of meters. We will see
that in the semiclassical regime the only WT compatible with the
positions of the pointers are the classical paths.

Let $\left\vert \psi (t_{i})\right\rangle $ be the initial state of
a dynamical system whose evolution is governed by a (possibly
time-dependent) Hamiltonian $H(t)$.\ Let us introduce  a meter
consisting of a particle positioned at $\mathbf{R}_{\kappa}^{0}$.
Its spatial wavefunction  $\left\langle
\mathbf{R}_{\kappa}\right\vert \left. \phi_{\kappa}\right\rangle $,
assumed to be tightly localized around $\mathbf{R}_{\kappa}^{0}$
acts as pointer. For convenience the wavefunction can be taken to be
a Gaussian,
 $\left\langle
\mathbf{R}_{\kappa}\right\vert \left. \phi_{\kappa}\right\rangle=
(2/\pi \Delta ^{2})^{1/2}e^{-\left( \mathbf{R}_{\kappa}-\mathbf{R}%
_{\kappa}^{0}\right) ^{2}/\Delta ^{2}}$ (we work from now on in a 2D
configuration space and use atomic units throughout).\ The local
coupling between the meter and the system is assumed to take place
during a small time interval $\tau$, triggered when the system and
pointer wavefunctions overlap.
The time-integrated interaction is taken as $I_{\kappa}=g\mathbf{r}\cdot \mathbf{R%
}_{\kappa}\theta ((4\Delta )^{2}-\left\vert
\mathbf{r}-\mathbf{R}_{\kappa}\right\vert ^{2})$, where $g$ is the
effective coupling strength and the last term is a unit-step
function accounting for the short range character of the interaction
(this term will be implicit in the rest of the paper). Assume now we
have a set of meters $\kappa=1,...,n$ positioned at
$\mathbf{R}_{\kappa}^{0}$
. Let $t_{\kappa}$ denote the mean interaction time of the
$\kappa$th pointer with the system. The initial state of the system
and meters $\left\vert \Psi (t_{i})\right\rangle =\left\vert \psi
(t_{i})\right\rangle \prod_{\kappa=1}^{n}\left\vert \phi
_{\kappa}\right\rangle $ evolves at time $t_{f}$ to \cite{fnt}
\begin{equation}
\left\vert \Psi (t_{f})\right\rangle
=U(t_{f},t_{n})e^{-iI_{n}}U(t_{n},t_{n-1})...e^{-iI_{1}}U(t_{1},t_{i})%
\left\vert\Psi (t_{i})\right\rangle%
\label{5}
\end{equation}%
where $U(t_{k+1},t_{k})$ denotes the unitary self evolution of the
system between two interactions; $k$ relabels the meters according
to the order in which they interact with the system.

At time $t_{f}$ a standard projective measurement is made in order to
postselect the system to a desired final state $\left\vert \chi
(t_{f})\right\rangle $. Expanding each $I_{k}$ in Eq. (\ref{5}) to first
order in the coupling $g$ leads to
\begin{align}
{\prod\limits_{k=1}^{n}}& \left\langle \mathbf{R}_{k}\right\vert
\left\langle \chi (t_{f})\right\vert \left. \Psi (t_{f})\right\rangle \simeq
\left\langle \chi (t_{f})\right\vert \left. \psi (t_{f})\right\rangle
\notag \\
& {\prod\limits_{k=1}^{n}}\exp \left[ -ig\left\langle \mathbf{r}%
(t_{k})\right\rangle _{W}\cdot \mathbf{R}_{k}\right] \phi _{k}(\mathbf{R}%
_{k},\mathbf{R}_{k}^{0})  \label{10}
\end{align}%
where $\left\langle \mathbf{r}(t_{k})\right\rangle _{W}$ is the weak value
[Eq. (\ref{2})] given here by
\begin{equation}
\left\langle \mathbf{r}(t_{k})\right\rangle _{W}\equiv \frac{\left\langle
\chi (t_{k})\right\vert \mathbf{r}\left\vert \psi (t_{k})\right\rangle }{%
\left\langle \chi (t_{k})\right\vert \left. \psi (t_{k})\right\rangle }
\label{12}
\end{equation}%
with $\mathbf{r}=x\mathbf{\hat{x}}+y\mathbf{\hat{y}.}$ Eqs. (\ref{10})-(\ref%
{12}) indicate that as a result of the interaction that took place at $t_{k}$%
, each meter wavefunction $\phi _{k}(\mathbf{R}_{k},\mathbf{R}_{k}^{0})$
will incur a phase-shift given by the WV $\left\langle \mathbf{r}%
(t_{k})\right\rangle _{W}$. As in the standard WM scenario
\cite{aav}, this phase-shift appears as a shift in the momentum
space wavefunction of each pointer. Since Eq. (\ref{10}) holds
provided $g$ and $\Delta $ are very small, the momentum space
wavefunctions will be broad, meaning a high number of events must be
recorded in order to observe each shift.

\begin{figure}[tb]
\includegraphics[height=8cm]{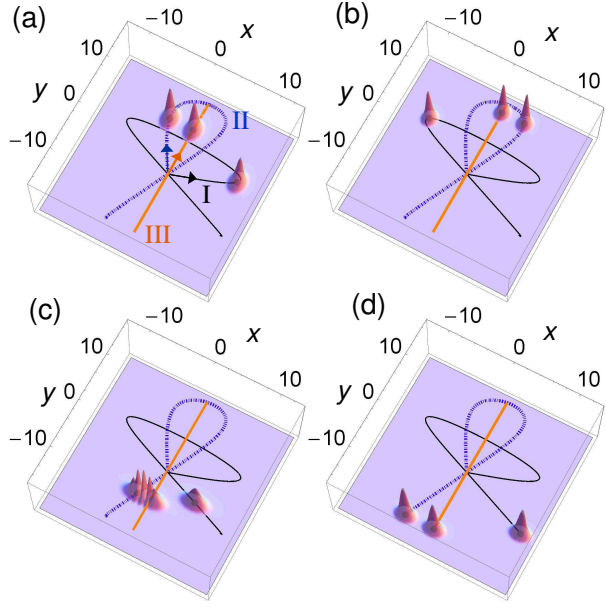}
\caption{Time evolution of the wavefunction initially ($t_{i}=0$) given by
Eq. (\ref{20}) with $\mathbf{r}_{0}=\mathbf{0}$ and the initial mean
momenta $\mathbf{p}_{j}, j=I,I\negmedspace I,I\negmedspace I\negmedspace I$
taken as indicated by the arrows in panel (a). The reference classical
trajectories $I$, $I\negmedspace I$ and $I\negmedspace I\negmedspace I$ are
shown resp. in black, dashed blue, and orange. (a) shows the wavefunction at
$t_{1}=0.7$, (b) at $t_{2}=2$, (c) at $t_{3}=3.15$ (after the wavepackets
cross the origin) and (d) at $t_{f}=3.65$, the time at which postselection
is made
. }
\label{fig1}
\end{figure}

The structure of $\left\langle \mathbf{r}(t_{k})\right\rangle _{W}$ deserves
a special comment. Each $\left\langle \mathbf{r}(t_{k})\right\rangle _{W}$
is defined at $t=t_{k}$ with an effective preselected state $\left\vert \psi
(t_{k})\right\rangle =U(t_{k},t_{i})\left\vert \psi (t_{i})\right\rangle $
being the initial state propagated \emph{forward} in time and an effective
postselected state $\left\langle \chi (t_{k})\right\vert =\left\langle \chi
(t_{f})\right\vert U(t_{f},t_{k})$ being the postselected state evolved
\emph{backward} in time; this property illustrates the close relation
between WM and time-symmetric formulations of quantum mechanics \cite{aav2}%
.\ Note that contrary to the usual definition of weak values, the effective
pre and postselected states defining $\left\langle \mathbf{r}%
(t_{k})\right\rangle _{W}$ cannot be chosen: only the initial and
the final states can be freely set. A WV at some intermediate time
$t_{k}$ reflects the interaction $I_{k}$ with the $k$th meter given
the unitary evolution of the preselected and postselected states of
the system. We can therefore envisage the set $\{t_{k},\left\langle
\mathbf{r}(t_{k})\right\rangle _{W}\}$ as defining a \emph{weak
trajectory} of the system evolving from an initial state to a final
postselected state as recorded by the pointers positioned at
$\mathbf{R}_{k}^{0},$ $k=1,...,n$.

For an arbitrary quantum system a WT will typically reflect the space-time
correlation between the forward evolution of the preselected state and the
backward evolution of the postselected state at the positions $\mathbf{R}%
_{k}^{0}$ of the weakly interacting meters. Although obtaining this
type of information is certainly of interest in general quantum
systems, the notion of weak trajectories is particularly suited to
investigate the evolution of a quantum system in the semiclassical
regime. In this regime a typical wavefunction evolves according to
the asymptotic form of the path integral
propagator \cite{schulman},%
\begin{align}
\psi(\mathbf{r},t) & =\int d\mathbf{r}^{\prime}\left\{ \sum_{cl}\frac {1}{%
\left( 2i\pi\hslash\right) }\left\vert \det\frac{\partial^{2}S_{cl}(\mathbf{r%
},\mathbf{r}^{\prime},t)}{\partial\mathbf{r\partial r}^{\prime}}\right\vert
^{1/2}\right.  \notag \\
& \left. \exp\left( iS_{cl}(\mathbf{r},\mathbf{r}^{\prime},t)/\hslash
-i\mu_{cl}\right) \right\} \psi(\mathbf{r}^{\prime},0)  \label{n10}
\end{align}
where $cl$ runs on the classical trajectories connecting $\mathbf{r}%
^{\prime} $ to $\mathbf{r}$ in time $t$ (from now on we set $t_{i}=0$) and
the term between $\left\{ ...\right\} $ is the semiclassical propagator
obtained from the asymptotics $(S_{cl}\gg\hslash)$ of the path integral form
of the evolution operator $U(t,0)$. $S_{cl}$ is the classical action and $%
\mu_{cl}$ the topological index of each path. Working out the full
semiclassical propagation is often a formidable task, especially as the
number of trajectories proliferate in the regimes where the semiclassical
approximation holds. However if the initial state is well localized, the
semiclassical propagation can be simplified by linearizing the action around
an initial and a final reference point linked in time $t$ by a central
classical trajectory, the guiding trajectory \cite{heller}. Linearization is
particularly relevant if $\psi(\mathbf{r}^{\prime},0)$ is chosen to be a
localized Gaussian%
\begin{equation}
\psi_{\mathbf{r}_{0},\mathbf{p}_{0}}(\mathbf{r}^{\prime},0)=\left( \frac{2}{%
\pi \delta^{2}}\right) ^{1/2}e^{-\left( \mathbf{r}^{\prime}-\mathbf{r}%
_{0}\right) ^{2}/\delta^{2}}e^{i\mathbf{p}_{0}\cdot\left( \mathbf{r}%
^{\prime}-\mathbf{r}_{0}\right) /\hslash}.  \label{14}
\end{equation}
The initial reference point is the maximum of the Gaussian, the linearized
action in Eq. (\ref{n10}) is a quadratic form, whereas the determinant
prefactor becomes a purely time-dependent term that can be written in terms
of the stability matrix of the guiding trajectory. This linearized
information along the central reference trajectory effectively replaces the
sum over $cl$.\

The integral (\ref{n10}) can then be performed exactly: the result
(often known as the Thawed Gaussian Approximation) is of the
form \cite{heller}%
\begin{align}
\psi_{\mathbf{r}_{0},\mathbf{p}_{0}}(\mathbf{r},t) & =\mathrm{Tr}\left[
\mathbf{A}(t)\right] e^{-\left( \mathbf{r}-\mathbf{q}(t)\right) \cdot\left[
\mathbf{M}(t)+i\mathbf{N}(t)\right] \cdot\left( \mathbf{r}-\mathbf{q}%
(t)\right) }  \notag \\
& e^{i\mathbf{p}(t)\cdot\left( \mathbf{r}-\mathbf{q}(t)\right) /\hslash
}e^{iS_{cl}(\mathbf{q}(t),\mathbf{r}_{0},t)/\hslash}  \label{15}
\end{align}
where $\left( \mathbf{q}(t),\mathbf{p}(t)\right) $ are the phase-space
coordinates of the guiding classical trajectory with initial conditions $(%
\mathbf{r}_{0},\mathbf{p}_{0})$ and $\mathbf{A,M}$ and $\mathbf{N}$ are
time-dependent matrices depending solely on the stability elements of the
guiding trajectory.\ The advantage of working in the linearized regime is
that by picking a preselected state of the form
\begin{equation}
\left\langle \mathbf{r}\right\vert \left. \psi(t_{i})\right\rangle =\sum
_{j}c_{j}\psi_{\mathbf{r}_{0},\mathbf{p}_{j}}(\mathbf{r},t_{i})  \label{20}
\end{equation}
i.e. a superposition of Gaussians (\ref{14}) launched in different
directions $\mathbf{p}_{j}$, one is dealing conceptually with the type of
problem defined by the semiclassical propagation (\ref{n10}) with a
simplified and perfectly controlled dynamics.\ The evolution operator $%
U(t_{1},t_{i})$ of Eq. (\ref{5}) propagates each term of Eq.
(\ref{20}) along the relevant guiding trajectory yielding at $t_{1}$
the superposition of evolved states $\sum_{j}c_{j}\psi_{\mathbf{r}%
_{0},\mathbf{p}_{j}}(\mathbf{r},t_{1})$, each state being given by
Eq. (\ref{15}) (recall that $\mathbf{q},\mathbf{p}$ and the matrices
$\mathbf{A,M}$ and $\mathbf{N}$ explicitly depend on $j$).

The postselected state will also be taken to be a Gaussian of the type (\ref%
{14}) localized in the vicinity of a chosen point $\mathbf{r}_{f}$ at time $%
t_{f}$%
\begin{equation}
\chi _{\mathbf{r}_{f},\mathbf{p}_{f}}(\mathbf{r},t_{f})=\left( \frac{2}{\pi
\delta _{f}^{2}}\right) ^{1/2}e^{-\left( \mathbf{r}-\mathbf{r}_{f}\right)
^{2}/\delta _{f}^{2}}e^{i\mathbf{p}_{f}\cdot \left( \mathbf{r}-\mathbf{r}%
_{f}\right) /\hslash }.  \label{25}
\end{equation}%
In the linearized approximation, finding the backward propagated state is
tantamount to obtaining the \emph{unique} solution of the form (\ref{15})
such that $\psi _{\mathbf{r}(t_{1}),\mathbf{p}(t_{1})}(\mathbf{r}%
,t_{f})=\chi _{\mathbf{r}_{f},\mathbf{p}_{f}}(\mathbf{r},t_{f})$: this gives
a wavefunction centered on a time-reversed classical trajectory having
boundary conditions $\left( \mathbf{r}_{f},\mathbf{p}_{f}\right) $ at $%
t=t_{f}$ and position $\mathbf{q}_f(t_k)$ at $t=t_k$.
Assume the meters lie at positions $%
\mathbf{R}_{k}^{0}$ where the overlap between the different branches
of the system wavefunction (\ref{20}) is negligible. The weak values
(\ref{12}) can then be computed exactly: if $\psi
_{\mathbf{r}_{0},\mathbf{p}_{j}}(\mathbf{r},t_{k}) $ (for all $j$)
or $\chi _{\mathbf{r}_{f},\mathbf{p}_{f}}(\mathbf{r},t_{k})$ vanish
in the vicinity of $\mathbf{R}_{k}^{0}$ the meter does not move:
there is no weak trajectory in the neighborhood of this point.\
Otherwise $\chi _{\mathbf{r}_{f},\mathbf{p}_{f}}(\mathbf{r},t_{k})$
overlaps at most with one branch, say $\psi
_{\mathbf{r}_{0},\mathbf{p}_{J}}$; denoting the distance between the
maxima of the wavepacket and of the postselected state at $t=t_{k}$
by
 $\boldsymbol{\epsilon }_{k}=\mathbf{q}%
_{J}(t_{k})-\mathbf{q}_{f}(t_{k}),$ the WV $\left\langle \mathbf{r}%
(t_{k})\right\rangle _{W}=\left\langle x(t_{k})\right\rangle _{W}\mathbf{%
\hat{x}}+\left\langle y(t_{k})\right\rangle _{W}\mathbf{\hat{y}}$ takes the
form%
\begin{align}
& \left\langle x(t_{k})\right\rangle _{W}=\left[ \mathbf{q}_{J}(t_{k})\cdot
\mathbf{\hat{x}+}a_{x}\boldsymbol{\epsilon }_{k}\cdot \mathbf{\hat{x}}%
+b_{x}\left( \mathbf{p}_{J}(t_{k})-\mathbf{p}_{f}(t_{k})\right) \cdot
\mathbf{\hat{x}}\right]   \notag \\
& +i\left[ g_{x}(t_{k})\boldsymbol{\epsilon }_{k}\cdot \mathbf{\hat{x}+}%
h_{x}\left( \mathbf{p}_{J}(t_{k})-\mathbf{p}_{f}(t_{k})\right) \cdot \mathbf{%
\hat{x}}\right]   \label{31}
\end{align}%
and analog expressions for $\left\langle y(t_{k})\right\rangle _{W}$.\ $a,b,g
$ and $h$ are time-dependent functions whose explicit forms are cumbersome
though straightforward to evaluate.

\begin{figure}[tb]
\includegraphics[height=4cm]{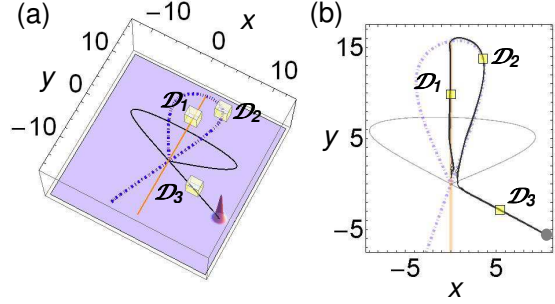}
\caption{(a) The postselected wavefunction, a Gaussian localized on
the guiding trajectory $I$, is shown, along with the positions of
the meters $\mathcal{D}_{1,2,3}$. Only the pointer $\mathcal{D}_{3}$
is affected, while $\mathcal{D}_{1,2}$ remain still: there is no WT
joining the corresponding positions. (b) The `average trajectories'
obtained from WM of the \emph{momentum} are plotted in solid black:
9 trajectories having their final positions on and near the maximum
of the postselected state are shown, along with the reference
trajectories (in faded colors). } \label{fig2}
\end{figure}

The structure of Eq. (\ref{31}) emphasizes the special r\^{o}le of the
postselected state with $(\mathbf{r}_{f},\mathbf{p}_{f})$ chosen such that
the backward evolved trajectory simply retraces the guiding trajectory $%
\mathbf{q}_{J}(t)$.\ For this special choice $\boldsymbol{\epsilon }_{k}=0$
and $\mathbf{p}_{J}(t_{k})=\mathbf{p}_{f}(t_{k})\ $for any value of $t_{k}$.
If $n$ meters happen to be in regions where $\chi $ and $\psi _{J}$ overlap
\emph{each} of these meters will record $\left\langle \mathbf{r}%
(t_{k})\right\rangle _{W}=\mathbf{q}_{J}(t_{k})$, ie the position of the
underlying classical trajectory.\ The WT $\left\{ t_{k},\left\langle \mathbf{%
r}(t_{k})\right\rangle _{W}\right\} $ thus corresponds to the
guiding trajectory of the linearized Feynman propagator. For any
other choice of postselection $\left\langle
\mathbf{r}(t_{k})\right\rangle _{W}$ (where defined) will yield a
complex number, with the real part indicating a registered position
that will be markedly different from the average position of the
wavepacket.\ The imaginary part of Eq. (\ref{31}) does not inform on
the value of the weakly measured observable (as is the rule for any
WV) but is related to the average backaction induced on the weak
meter by the postselection \cite{disturb}.

For the purpose of illustration -- and to avoid spurious effects due
to the quality of the linearized approximation -- we will take a 2D
time-dependent linear oscillator (TDLO).\ The linearized propagator
for the TDLO is quantum mechanically exact, while the varying
amplitudes capture many features of semiclassical systems with more
involved dynamics. The TDLO is often employed to model diverse
systems, like ions in a trap \cite{tdlo}. The
Hamiltonian for the system is $%
H=(P_{x}^{2}+P_{y}^{2})/2m+mV_{x}(t)x^{2}+mV_{y}(t)y^{2}$ where for
definiteness we choose $V_{i}(t)=\xi _{i}-\upsilon _{i}\cos \left(
2\omega _{i}t\right) $ ($i=x,y$; $\xi ,\upsilon $ and $\omega $ are
constants). The wavefunction (\ref{15}) is obtained directly by
employing standard path integral techniques \cite{schulman}; the
classical trajectories can be found in closed form from the
solutions of Ermakov systems \cite{ermakov}. The preselected state
(\ref{20}) is taken as the superposition of 3 Gaussians at the
origin with mean momenta as shown in Fig. 1(a). The maximum of each
wavepacket then evolves by following the guiding trajectory, $I$,
$I\negmedspace I$ or $I\negmedspace I\negmedspace I$ shown in Fig.\
1.

Let us first set the postselected state (\ref{25}) with $\mathbf{r}_{f}=%
\mathbf{q}_{I}(t_{f})$ and $\mathbf{p}_{f}=\mathbf{p}_{I}(t_{f})$
and let us position the meters $\mathcal{D}_{k}$ as shown in Fig.\
2(a). The backward evolution of $\left\vert \chi(t_{f})\right\rangle
$ simply retraces trajectory $I$ backwards. Therefore the pointer in
$\mathcal{D}_{3}$ displays
according to Eq. (\ref{31}) the position $\mathbf{q}_{I}(t_3)$ while $%
\mathcal{D}_{2}$ and $\mathcal{D}_{1}$ do not move at all (no overlap with $%
\left\vert \chi(t)\right\rangle $ at any $t$). One concludes that
the `particle' went through $\mathcal{D}_{3}$ but not through
$\mathcal{D}_{1}$ and $\mathcal{D}_{2}$. If instead of
$\mathcal{D}_{1}$ and $\mathcal{D}_{2}$ other meters
$\mathcal{D}_{1}^{\prime}$ and $\mathcal{D}_{2}^{\prime}$ positioned
as shown in Fig.\ 3(a) are employed, then these pointers display
respectively the WV $\mathbf{q}_{I}(t_1)$ and $\mathbf{q}_{I}(t_2)$:
the `particle' went through
$\mathcal{D}_{1}^{\prime}$,$\mathcal{D}_{2}^{\prime}$ and
$\mathcal{D}_{3}$. Hence one concludes (possibly by inserting
additional devices) that the `particle' took the WT defined by the
classical trajectory $I$. Note that according to Eq. (\ref{31})
there is no quantum state of the form (\ref{25})
that can yield a WT going through $\mathcal{D}_{1}$, $\mathcal{D}_{2}$ and $%
\mathcal{D}_{3}$. This is due to the fact, implied by the propagator (\ref%
{n10}), that there does not exist a wavepacket arriving in the neighborhood
of $\mathbf{r}_{f}$ at time $t_{f}$ that would have previously visited the
neighborhoods of $\mathcal{D}_{1}$, $\mathcal{D}_{2}$ and $\mathcal{D}_{3}$.

The last remark highlights the incompatibility between the `weak
trajectories' defined here by consecutive WM of the position and the
`average trajectories' (AT) defined by a WM of the \emph{momentum}
immediately postselected to a given position. By repeating these weak
momentum measurements for different postselected positions, a velocity field
is obtained.\ The AT are precisely the trajectories built on this velocity
field. They have been experimentally observed recently for photons in a
double slit setup \cite{at}.\ It was previously known \cite{dbb} that their
dynamics is governed by the law of motion of the de Broglie-Bohm theory \cite%
{bohm}, i.e. by the probability flow, whereas the WT are generated
by the semiclassical propagator (\ref{n10}). The mismatch
\cite{mismatch} between de Broglie-Bohm and classical trajectories
in semiclassical systems hinges on the fact that when wavepackets
interfere, the overall mean velocity field differs from the group
velocity of each individual wavepacket. The mismatch is illustrated
here in Fig.\ 2(b): we have computed numerically \cite{fnt}
several AT arriving in the neighborhood of $\mathbf{r}_{f}=\mathbf{q}%
_{I}(t_{f})$.\ These AT go indeed through $\mathcal{D}_{1}$, $\mathcal{D}%
_{2} $ and $\mathcal{D}_{3}$: starting near the origin, they first move in
the vicinity of the guiding trajectory $I\negmedspace I\negmedspace I$, then
travel along trajectory $I\negmedspace I$ and thereafter `jump' so as to
move along trajectory $I$.

\begin{figure}[tb]
\includegraphics[height=4cm]{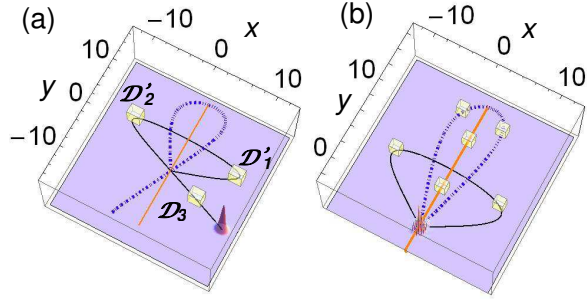}
\caption{(a): The postselected wavefunction (the same as displayed in Fig.
2) is shown with the meters $\mathcal{D}^{\prime}_{1},\mathcal{D}%
^{\prime}_{2}$ and $\mathcal{D}_{3}$. Each of these pointers
indicates a weak value $\mathbf{q}_{I}(t_k)$ : the evolution of the
system along the reference trajectory $I$ has been measured weakly.
(b): Postselection now takes places at $t_{O}=2.84$ when the
wavepackets return simultaneously to the origin. The postselected
state (defined in the text) is plotted along with the pointers
positioned along the reference trajectories $I,I\negmedspace I,I\negmedspace %
I\negmedspace I.$ \emph{All} the meters yield weak values in
agreement with their position along the relevant classical
trajectory : the semiclassical sum over paths formulation has thus
been measured weakly. } \label{fig3}
\end{figure}

Finally, consider choosing postselection at $t_f=t_{O}$ when trajectories $%
I,I\negmedspace I$ and $I\negmedspace I\negmedspace I$ first return to the
origin, with the postselected state chosen as the superposition $\chi _{O}(%
\mathbf{r},t_{O})=\sum_{j}\chi_{\mathbf{r}_{f}=0,\mathbf{p}_{Oj}}(\mathbf{r}%
,t_{O})$, with $\chi_{\mathbf{r}_{f}=0,\mathbf{p}_{Oj}}$ given by Eq. (\ref%
{25}) and $\mathbf{p}_{Oj}=\mathbf{p}_{j}(t_{O})$, $j=I,I\negmedspace I,I%
\negmedspace I\negmedspace I.$ Several pointers are positioned as
shown in Fig. 3(b). By construction the backward evolution of
$\chi_{O}$
yields a superposition of wavepackets retracing trajectories $I,I%
\negmedspace I$ and $I\negmedspace I\negmedspace I$ respectively.
Therefore \emph{all} the pointers will display a WV consistent with
their position along one of the three trajectories, indicating the
`particle' was there.\ This is an experimentally realizable way to
catch the essence of the path integral approach in the semiclassical
regime: weakly interacting meters indicate the `particle' takes
\emph{simultaneously} all the available classical paths. In contrast
a strong projective measurement would of course yield a definite
outcome on only \emph{one} of the paths.

To sum up, we have defined `weak trajectories' allowing to observe the paths
taken by a quantum system in the semiclassical regime by direct weak
measurements of the position. A consequence worth exploring concerns the
possibility of employing this scheme to reconstruct the unknown propagator
of a semiclassical system from the observed WT obtained from a grid of weak
detectors while filtering postselected states. Possible experimental
realizations could be considered in systems in which wavepackets with a
controlled dynamics can be engineered \cite{wavepackets}. The present setup
may also be used in designing pre-post selected quantum paradoxes containing
dynamical ingredients.

\appendix
\section{Derivation of the evolution equations, Eqs. (2)-(4)}

We clarify in this Appendix the assumptions and simplifications
leading to the evolution Eqs. (2)-(4).\

\subsection{Single meter}

Let $U(t,t^{\prime})$ denote the evolution operator for the system. $U$ obeys%
\begin{equation}
i\hslash\partial_{t}U(t,t^{\prime})=H(t)U(t,t^{\prime})
\end{equation}
where $H(t)=\mathbf{p}^{2}/2m+V(\mathbf{r},t)$ is the system
Hamiltonian. Let us introduce a meter consisting of a mass
positioned at $\mathbf{R}_{\kappa}^{0}$, that can move freely in the
2D plane. Its spatial wavefunction  $\left\langle
\mathbf{R}_{\kappa}\right\vert \left. \phi_{\kappa}\right\rangle $,
assumed to be tightly localized around the central position
$\mathbf{R}_{\kappa}^{0}$, acts as a pointer; for definiteness we
can choose a Gaussian wavefunction
$\phi_{\kappa}(\mathbf{R}_{\kappa})=(2/\pi\Delta^{2})^{1/2}e^{-\left(
\mathbf{R}_{\kappa}-\mathbf{R}_{\kappa}^{0}\right)
^{2}/\Delta^{2}}$. We represent the interaction between the system
and the meter by the interaction Hamiltonian
\begin{equation}
H_{\kappa}(t)=\gamma(t)\mathbf{r}\cdot\mathbf{R}_{\kappa}\theta((4\Delta)^{2}-\left\vert
\mathbf{r}-\mathbf{R}_{\kappa}\right\vert ^{2}). \label{hk}%
\end{equation}
$\theta$ is the unit-step function with its argument indicating the
short-range character of the interaction.\ $\gamma(t)$ is a smooth
even function of
$t$ (symmetric about $t_{\kappa}$) vanishing outside an interval $t_{\kappa}%
-\tau/2<t<t_{\kappa}+\tau/2$ and obeying $\int_{t_{\kappa}-\tau/2}^{t_{\kappa}+\tau/2}%
\gamma(t)dt=\hslash g$; $\tau$ appears as the duration of the
measurement, which will be taken to be short relative to the
timescale of the system dynamics.\ $t_{\kappa}$ is the mean time at
which the interaction takes place. The interaction is turned on when
the system and pointer wavefunctions overlap significantly (ie,
above some threshold). The interaction evolution operator between
$t=t_{\kappa}-\tau/2$ and $t_{\kappa}+\tau/2$ is thus given by
\footnote{Note that
there is no time-ordering problem since $H_\kappa$ commutes with itself at different times.}%
\begin{equation}
U_{\kappa}(t_{\kappa}+\tau/2,t_{\kappa}-\tau/2)=\exp\left(
-\frac{i}{\hslash}\int
_{t_{\kappa}-\tau/2}^{t_{\kappa}+\tau/2}H_{\kappa}(t^{\prime})dt^{\prime}\right)
\equiv
\exp-iI_{\kappa} \label{defi}%
\end{equation}
with
$I_{\kappa}=g\mathbf{r}\cdot\mathbf{R}_{\kappa}\theta((4\Delta)^{2}-\left\vert
\mathbf{r}-\mathbf{R}_{\kappa}\right\vert ^{2}).\ U_{\kappa}$ obeys%
\begin{equation}
i\hslash\partial_{t}U_{\kappa}(t,t^{\prime})=H_{\kappa}(t)U_{\kappa}(t,t^{\prime}).
\end{equation}
We suppose this type of meter, can only be triggered once (it cannot
interact again with the system once it has been triggered).

Assume the system wavefunction $\psi(\mathbf{r},t)$ enters the
region near $\mathbf{R}_{\kappa}^{0}$ triggering the interaction.
This happens (by definition) at $t=t_{\kappa}-\tau/2.$ The evolution
up to $t=t_{\kappa}+\tau/2$ is then generated by
the total evolution operator $U_{T}(t,t_{\kappa}-\tau/2)$ obtained formally from%
\begin{equation}
i\hslash\partial_{t}U_{T}(t,t_{\kappa}-\tau/2)=\left[
H(t)+H_{\kappa}(t)\right] U_{T}(t,t_{\kappa}-\tau/2).
\end{equation}
Since the interaction is weak, $H_{\kappa}(t)$ appears as a small
perturbation and $U_{T}$ is best obtained in the interaction
picture. Let $\left\vert
\Psi(t_{\kappa}-\frac{\tau}{2})\right\rangle \equiv\left\vert
\psi(t_{\kappa}-\frac {\tau}{2})\right\rangle \left\vert
\phi_{\kappa}(t_{\kappa}-\frac{\tau}{2})\right\rangle $ denote the
total wavefunction when the interaction is turned on. Then by
definition of the interaction picture (eg \cite{louisell}) one has
\begin{equation}
\left\vert \Psi(t_{\kappa}+\frac{\tau}{2})\right\rangle =\tilde{U}_{T}(t_{\kappa}%
+\frac{\tau}{2},t_{\kappa}-\frac{\tau}{2})\left\vert \Psi(t_{\kappa}-\frac{\tau}%
{2})\right\rangle \label{psitot}%
\end{equation}
where%
\begin{equation}
\tilde{U}_{T}(t_{\kappa}+\frac{\tau}{2},t_{\kappa}-\frac{\tau}{2})\equiv U(t_{\kappa}%
+\frac{\tau}{2},t_{\kappa}-\frac{\tau}{2})\tilde{U}_{\kappa}(t_{\kappa}+\frac{\tau}{2}%
,t_{\kappa}-\frac{\tau}{2}). \label{totev}%
\end{equation}
$\tilde{U}_{\kappa}$ is the interaction evolution operator in the
interaction picture, formally
given by the integral equation%
\begin{equation}
\tilde{U}_{\kappa}(t_{\kappa}+\frac{\tau}{2},t_{\kappa}-\frac{\tau}{2})=I-\frac{i}{\hslash
}\int_{t_{\kappa}-\frac{\tau}{2}}^{t_{\kappa}+\frac{\tau}{2}}\tilde{H}_{\kappa}(t^{\prime
})\tilde{U}_{\kappa}(t^{\prime},t_{\kappa}-\frac{\tau}{2})dt^{\prime} \label{integ}%
\end{equation}
where $\tilde{H}_{\kappa}(t^{\prime})$ is the interaction
Hamiltonian in the
interaction picture%
\begin{equation}
\tilde{H}_{\kappa}(t)=U^{-1}(t,t_{\kappa}-\frac{\tau}{2})H_{\kappa}(t)U(t,t_{\kappa}-\frac{\tau
}{2}). \label{intph}%
\end{equation}
To first order in the interaction Eq. (\ref{integ}) yields%
\begin{equation}
\tilde{U}_{\kappa}(t_{\kappa}+\frac{\tau}{2},t_{\kappa}-\frac{\tau}{2})=I-\frac{i}{\hslash
}\int_{t_{\kappa}-\frac{\tau}{2}}^{t_{\kappa}+\frac{\tau}{2}}\tilde{H}_{\kappa}(t^{\prime
})dt^{\prime}. \label{fo}%
\end{equation}

The computation of Eq. (\ref{fo}) can be simplified by assuming that
the duration of the interaction $\tau$ is small relative to the
system evolution.\ The Taylor expansion of $U(t),$ a standard tool
when determining the evolution operator in time dependent problems
\cite{chap} can then be simplified
by taking the sole zero order term $U(t,t_{\kappa}-\tau/2)\approx U(t_{\kappa},t_{\kappa}%
-\tau/2)$.\ Using this approximation in Eq. (\ref{intph}) with
$t=t_\kappa+\tau/2$ Eq. (\ref{fo})
becomes%
\begin{equation}
\tilde{U}_{\kappa}(t_{\kappa}+\frac{\tau}{2},t_{\kappa}-\frac{\tau}{2})\approx U^{-1}%
(t_{\kappa},t_{\kappa}-\frac{\tau}{2})\exp\left(  -iI_{\kappa}\right)  U(t_{\kappa},t_{\kappa}%
-\frac{\tau}{2}).
\end{equation}
where we have identified the first order of the exponential with the
exact operator (since the interaction is weak) and used Eq.
(\ref{defi}).\ The
wavefunction (\ref{psitot}) after the interaction takes the form%
\begin{align}
\left\vert \Psi(t_{\kappa}+\frac{\tau}{2})\right\rangle  &  \approx U(t_{\kappa}%
+\frac{\tau}{2},t_{\kappa})\exp\left(  -iI_{\kappa}\right)  U(t_{\kappa},t_{\kappa}-\frac{\tau}%
{2})\left\vert \Psi(t_{\kappa}-\frac{\tau}{2})\right\rangle \label{ev1}\\
&  \approx U(t_{\kappa}+\frac{\tau}{2},t_{\kappa})\exp\left(
-iI_{\kappa}\right)  \left\vert
\psi(t_{\kappa})\right\rangle \left\vert \phi_{\kappa}\right\rangle \label{ev2}%
\end{align}
whose physical meaning is transparent: the time-dependent
interaction that takes place between
$t_{\kappa}-\tau/2<t<t_{\kappa}+\tau/2$ appears as an effective
interaction applied at $t=t_{\kappa}$ (the mean interaction time),
the system evolving "freely" between $t_{\kappa}-\tau/2$ and
$t_{\kappa}$ and again between $t_{\kappa}$ and $t_{\kappa}+\tau/2.$
For times $t>t_{\kappa}+\tau/2$ the system is of course subjected to
its own self-evolution $U(t_{f},t_{\kappa}+\tau/2)$ until a standard
projective measurement is made at $t=t_{f}$ to post-select to a
final state $\left\vert \chi(t_{f})\right\rangle$. Since
\begin{equation}
\left\langle \chi(t_{f})\right\vert
U(t_{f},t_{\kappa}+\tau/2)U(t_{\kappa}+\frac{\tau
}{2},t_{\kappa})\exp\left( -iI_{\kappa}\right)  \left\vert
\psi(t_{\kappa})\right\rangle =\left\langle
\chi(t_{\kappa})\right\vert \exp\left(  -iI_{\kappa}\right)
\left\vert
\psi(t_{\kappa})\right\rangle \label{fin1}%
\end{equation}
the postselected state can be said to evolve backward up to the
effective interaction time $t_{\kappa}.\ $At this point, the
procedures usually employed in the weak measurement litterature are
applied to the right handside of Eq. (\ref{fin1}) to derive the weak
value.

\subsection{Several meters}

We now extend the formulas derived above for a single meter to the
case involving several pointers.\ In order to obtain a simple
compact formula, it is convenient to work with a time-ordered
sequence of interacting pointers. The physical picture one should
then have in mind is that of a single localized wavepacket
interacting successively with several pointers (this is how
trajectories can be defined operationally). Indeed, if the
wavefunction is widely extended in space, it can in principle
interact simultaneously with several meters and the global evolution
operator cannot be simplified. If the wavefunction has different
branches, like a superposition of localized wavepackets, then each
branch gives rise to its own sequence \footnote{Recall that as
mentioned in the paper, we are disregarding the meters that would be
positioned in regions where different branches overlap -- the
behavior of such meters are not covered by the present treatment.}.

Assume we have overall $N$ meters.\ As above the spatial
wavefunction of each meter acts as a pointer, and the system-meter
interaction is encapsulated in the Hamiltonian $H_{\kappa}$ given by
Eq. (\ref{hk}) where $\kappa=1,...,N$ labels the $\kappa$th meter as
determined by its position. Although the way in which the $N$ meters
are positioned is not particularly important, it makes sense to
envisage they are positioned in a grid. Then only a handful $n$ of
these $N$ meters will interact with the system wavepacket. The order
in which the meters will interact depends on the wavepacket initial
position and dynamics. We relabel the meters with the index $k,$
reflecting the order of interaction with the system ($k=1$
corresponds to the meter interacting first with the system, $k=2$ to
the second meter having interacted with the system etc.,); the
correspondence between $\kappa$ (which is fixed) and $k$ depends
obviously on the system dynamics. This relabeling can be done
beforehand if the system dynamics is known, or retrospectively if
the system dynamics is to be inferred.

At any rate, once the relabeling is done, the initial state
$\left\vert \Psi(t_{i})\right\rangle =\left\vert
\psi(t_{i})\right\rangle \prod_{\kappa =1}^{N}\left\vert
\phi_{\kappa}\right\rangle $ can be rewritten (disregarding
the $N-n$ meters that will not be interacting with the wavefunction) as%
\begin{equation}
\left\vert \Psi(t_{i})\right\rangle =\left\vert
\psi(t_{i})\right\rangle \prod_{k=1}^{n}\left\vert
\phi_{k}\right\rangle .
\end{equation}
Let us denote, similarly to the single meter case, the mean
interaction time of the system with the first ($k=1$) meter by
$t_{1}.$ The system evolves from $t_{i}$ to $t_{1}-\tau/2$ according
to its own self-evolution
$U(t_{1}-\tau/2,t_{i})$, and then the interaction takes place for $t_{1}%
-\tau/2<t<t_{1}+\tau/2$.\ Hence according to Eq. (\ref{ev2}), after
the
interaction with the first meter has ended the global state becomes%
\begin{equation}
\left\vert \Psi(t_{1}+\frac{\tau}{2})\right\rangle =U(t_{1}+\frac{\tau}%
{2},t_{1})\exp\left(  -iI_{1}\right)  U(t_{1},t_{i})\left\vert \psi
(t_{i})\right\rangle \left\vert \phi_{1}\right\rangle \prod_{k=2}%
^{n}\left\vert \phi_{k}\right\rangle .
\end{equation}
Right before interacting with the second ($k=2)$ meter, the
self-evolution of
the system evolves the quantum state to%
\begin{align}
\left\vert \Psi(t_{2}-\frac{\tau}{2})\right\rangle  &  =U(t_{2}-\frac{\tau}%
{2},t_{1}+\frac{\tau}{2})\left\vert \Psi(t_{1}+\frac{\tau}{2})\right\rangle \\
&  =U(t_{2}-\frac{\tau}{2},t_{1})\exp\left(  -iI_{1}\right)  U(t_{1}%
,t_{i})\left\vert \psi(t_{i})\right\rangle \prod_{k=1}^{n}\left\vert
\phi _{k}\right\rangle .
\end{align}
The interaction of the system with the second meter in the interval
$t_{2}-\tau/2<t<t_{2}+\tau/2$ is treated again by using Eqs. (\ref{ev1}%
)-(\ref{ev2}) with $k=2$ yielding%
\begin{align}
\left\vert \Psi(t_{2}+\frac{\tau}{2})\right\rangle  &  =U(t_{2}+\frac{\tau}%
{2},t_{2})\exp\left(  -iI_{2}\right)
U(t_{2},t_{2}-\frac{\tau}{2})\left\vert
\Psi(t_{2}-\frac{\tau}{2})\right\rangle \\
&  =U(t_{2}+\frac{\tau}{2},t_{2})\exp\left(  -iI_{2}\right)  U(t_{2}%
,t_{1})\exp\left(  -iI_{1}\right)  U(t_{1},t_{i})\left\vert \psi
(t_{i})\right\rangle \prod_{k=1}^{n}\left\vert \phi_{k}\right\rangle
.
\end{align}
The same procedure is followed for the subsequent interactions.\

After the system has interacted with the last meter the
self-evolution up to the post-selection time $t_{f}$ gives the
additional term $U(t_{f},t_{n}),$
from which Eq. (2) is obtained:%
\begin{equation}
\left\vert \Psi(t_{f})\right\rangle =U(t_{f},t_{n})e^{-iI_{n}}U(t_{n}%
,t_{n-1})...e^{-iI_{1}}U(t_{1},t_{i})\left\vert
\psi(t_{i})\right\rangle
\prod_{k=1}^{n}\left\vert \phi_{k}\right\rangle \mathbf{.} \label{fin}%
\end{equation}
Finally, each interaction exponential is expanded to first order in
$g$, and upon post-selection, the projection
${\prod\limits_{j=1}^{n}}\left\langle \mathbf{R}_{j}\right\vert
\left\langle \chi(t_{f})\right\vert \left. \Psi(t_{f})\right\rangle
$ becomes (keeping only the terms to first order in
$g$)%
\begin{equation}
{\prod\limits_{j=1}^{n}}\left\langle \mathbf{R}_{j}\right\vert
\left\langle \chi(t_{f})\right\vert \left.  \Psi(t_{f})\right\rangle
=\left\langle
\chi(t_{f})\right\vert U(t_{f},t_{i})-\sum_{k=1}^{n}iU(t_{f},t_{k}%
)I_{k}U(t_{k},t_{i})\left\vert \psi(t_{i})\right\rangle
\prod_{j=1}^{n}\phi_{j}(\mathbf{R}_{j}).
\end{equation}
Factorizing $\left\langle \chi(t_{f})\right\vert
U(t_{f},t_{i})\left\vert \psi(t_{i})\right\rangle $ and
re-exponentiating the first-order expression gives Eq. (3).\ This
completes the derivation of the evolution equation for the system
weakly coupled to a series of weak meters.

\section{Determination of the ``Average Trajectories"}
We briefly recall in this Appendix how ``Average Trajectories" arise
from weak measurements and detail how the trajectories plotted in
Fig. 2(b) were computed.

\subsection{Velocity field and weak measurements}

The standard well-known form of the quantum mechanical probability
current for
the system is%
\begin{equation}
\mathbf{j}(\mathbf{r},t)=\frac{i\hslash}{2m}\left[  \psi(\mathbf{r}%
,t)\nabla\psi^{\ast}(\mathbf{r},t)-\psi^{\ast}(\mathbf{r},t)\nabla
\psi(\mathbf{r},t)\right]  .\label{cur}%
\end{equation}
From the current density we can define (by analogy eg with classical
fluid mechanics) a velocity field through
\begin{equation}
\mathbf{v}(\mathbf{r},t)=\frac{\mathbf{j}(\mathbf{r},t)}{\left\vert
\psi(\mathbf{r},t)\right\vert ^{2}}.\label{vel}%
\end{equation}
This gives the local velocity along the streamlines of the
probability flow. In the de Broglie-Bohm interpretation of quantum
mechanics \cite{bohm}, where particles are assumed to move along
these streamlines, $\mathbf{v}(\mathbf{r},t)$ represents the
velocity of the Bohmian particle. Employing the polar decomposition
$\psi(\mathbf{r},t)\equiv\rho^{1/2}(\mathbf{r},t)\exp
(i\sigma(\mathbf{r},t)/\hslash),$ Eqs. (\ref{cur}) and (\ref{vel}) give%
\begin{equation}
\mathbf{v}(\mathbf{r},t)=\frac{\mathbf{\triangledown}\sigma(\mathbf{r},t)}%
{m}.\label{bv}%
\end{equation}

Consider now a weak measurement of the momentum immediately followed
by a projective position measurement, with a postselection to some
given state $\left\vert \chi\right\rangle \equiv\left\vert
\mathbf{r}\right\rangle .$ The corresponding weak value
$\left\langle \mathbf{p}\right\rangle _{W}$ is
obtained by applying Eq. (1), yielding%
\begin{align}
\left\langle \mathbf{p}\right\rangle _{W}  & =\frac{\left\langle
\chi\right\vert \mathbf{\hat{p}}\left\vert \psi\right\rangle
}{\left\langle
\chi\right\vert \left.  \psi\right\rangle }\\
& =m\mathbf{v}(\mathbf{r},t)-i\hslash\frac{\mathbf{\triangledown}%
\rho(\mathbf{r},t)}{2\rho(\mathbf{r},t)}.
\end{align}
The real part of the weak value, which is the part that is
experimentally measurable \cite{disturb}, is proportional to the
hydrodynamic velocity field defined by Eqs. (\ref{vel}) or
(\ref{bv}).

Note that  $\left\langle \mathbf{p}\right\rangle _{W}$ can also be
given an 'operational' derivation in terms of position measurements
\cite{dbb}: assume the position is weakly measured at time $t$ and
strongly measured immediately after at time $t+\varepsilon$ and
found at position $\mathbf{r}_{f}$.\ The weak value of the position
consistent with the post-selection at
$\mathbf{r}_{f}$ is given by%
\begin{equation}
\left\langle \mathbf{r}\right\rangle _{W}=\frac{\left\langle \mathbf{r}%
_{f}\right\vert e^{-iH\varepsilon/\hslash}\mathbf{\hat{r}}\left\vert
\psi\right\rangle }{\left\langle \mathbf{r}_{f}\right\vert
e^{-iH\varepsilon /\hslash}\left\vert \psi\right\rangle }.
\end{equation}
To first order in $\varepsilon\rightarrow0$ one obtains after some
manipulations
\begin{equation}
\frac{\left\langle \mathbf{p}\right\rangle
_{W}}{m}=\frac{1}{\varepsilon }\left(  \mathbf{r}_{f}-\left\langle
\mathbf{r}\right\rangle _{W}\right)
\label{av}%
\end{equation}
so that the velocity field appears as the real part of the
difference $\left( \mathbf{r}_{f}-\left\langle
\mathbf{r}\right\rangle _{W}\right)  /\varepsilon $. Since
experimentally $\left\langle \mathbf{r}\right\rangle _{W}$ is
obtained by averaging over several runs, the trajectories inferred
from the velocity field appear as ``average trajectories''.\ From a
formal standpoint $\left\langle \mathbf{r}\right\rangle _{W}$ is a
transition matrix element where the final position is perfectly well
localized but the ``initial'' position (at time $t$) is not defined
better than $\psi(\mathbf{r},t)$ is; hence Eq. (\ref{av}) involves
averaging over this spatial region. Another different sense in which
these trajectories can be said to be ``average'' is that the
velocity field (\ref{vel}) is defined from the \emph{net} flow: if
several waves propagating in different directions overlap,
$\mathbf{v}(\mathbf{r},t)$ is the velocity of the resulting flow
averaged from the flows produced by the individual waves. This is
what makes the average trajectories unclassical even in the
classical limit \cite{shps}.

\subsection{Computation of the ``average trajectories"}

Computing the average trajectories means computing the Bohmian
trajectories \cite{bohm}.\ The usual method involves 3 steps: (i)
Obtain the wavefunction of the system, here given by
$\psi(\mathbf{r},t)=\sum_{j}c_{j}\psi
_{\mathbf{r}_{0},\mathbf{p}_{j}}(\mathbf{r},t)$ where $\psi_{\mathbf{r}%
_{0},\mathbf{p}_{j}}(\mathbf{r},t)$ is computed from Eq. (7); (ii)
Compute the logarithmic derivatives
$\partial_{\alpha}\psi(\mathbf{r},t)/\psi (\mathbf{r},t)$ with
$\alpha=x,y$ whose imaginary part is proportional to
$\mathbf{v}(\mathbf{r},t)$; (iii) Integrate the equations%
\begin{equation}
\left\{
\begin{array}
[c]{c}%
\frac{dx}{dt}=\frac{\hslash}{m}\operatorname{Im}\frac{\partial_{x}\psi
(x,y,t)}{\psi(x,y,t)}\\
\frac{dy}{dt}=\frac{\hslash}{m}\operatorname{Im}\frac{\partial_{y}\psi
(x,y,t)}{\psi(x,y,t)}%
\end{array}
\right.  \label{dbt}%
\end{equation}
whose solution starting from a known initial position
$(x_{i},y_{i})$ is unique.

Here, in order to obtain the average trajectories shown in Fig.\
2(b) we have determined $\psi(\mathbf{r},t)$ for the TDLO from the
linearized propagator (which turns out to be the exact
quantum-mechanical propagator in this case). The boundary condition
was taken at the final position
$\mathbf{r}_{f}=\mathbf{q}_{I}(t_{f})$ (at $t=t_{f})$ and Eq.
(\ref{dbt}) was integrated backward in time \footnote{The exact
momenta are $\mathbf{p}_{I,I\negmedspace I,I \negmedspace
I\negmedspace I}=(17,7),(-7,15),(0,15)$ and the coefficients in Eq.
(8) are $c_{I,I\negmedspace I,I\negmedspace I\negmedspace
I}=0.32,0.35,0.33$ .}.\ 9 trajectories are shown in Fig.\ 2(b): the
one at position $\mathbf{r}_{f}=\mathbf{q}_{I}(t_{f})$ (which
corresponds to the maximum of the probability distribution) and 8\
other trajectories in the
neighborhood of that point, with final positions $(x_{f},y_{f})=(x_{I}%
(t_{f})\pm0.05,y_{I}(t_{f})\pm0.05)$ where the number $0.05$ is
given in the same units as those of the figure.


\begin{thebibliography}{99}
\bibitem{sc} M. Brack and R Bhaduri, \emph{Semiclassical physics} (Westview,
Boulder, 1997);

\bibitem{classquant} M. R. Haggerty et al, Phys. Rev. Lett. 81, 1592 (1998);
A. Matzkin et al, Phys. Rev. A 68, 061401(R) (2003); Z. Chen et al, Phys.
Rev. Lett. 102, 244103 (2009); J. D. Wright et al, Phys. Rev. A 81, 063409
(2010)

\bibitem{aav} Y. Aharonov et al Phys. Rev. Lett. 60, 1351 (1988).

\bibitem{aav2} Y. Aharonov et al Phys. Today 63, 11, 27 (2010).

\bibitem{wma} O. Hosten and P. Kwiat, Science 319, 787 (2008); P. Ben Dixon
et al Phys. Rev. Lett. 102, 173601 (2009).

\bibitem{wm} See eg Y. Aharonov and A. Botero Phys. Rev. A 72, 052111
(2005); G. Mitchison et al, Phys. Rev. A 76, 062105 (2007); J. S.
Lundeen and A. M. Steinberg, Phys. Rev. Lett. 102, 020404 (2009); A.
Tanaka Phys. Lett. A 297 307 (2002); Y. Kedem and L. Vaidman Phys.
Rev. Lett. 105, 230401 (2010); M. Iinuma et al, N. J. Phys. 13
033041 (2011); M.E. Goggin et al, Proc. Natl. Acad. Sci. U.S.A. 108,
1256 (2011).

\bibitem{at} Kocsis et al, Science 332, 1170 (2011).

\bibitem{fnt}See Appendix for details.


\bibitem{schulman} L. S. Schulman \emph{Techniques and Applications of Path
Integration} (Wiley, New York, 1996).

\bibitem{heller} E. J. Heller in \emph{Chaos and Quantum Physics} (Elsevier,
Amsterdam, 1991), Ch. 9.

\bibitem{disturb}A. M. Steinberg, Phys Rev Lett 74 2405 (1995); J. Dressel and A. N. Jordan, Phys. Rev. A 85 012107 (2012).

\bibitem{tdlo}N. Menicucci and G. J. Milburn, Phys Rev A 76 052105 (2007).

\bibitem{ermakov} S.V. Lawande and A.K. Dhara, Phys. Rev. A 30 560 (1984).


\bibitem{dbb} H. Wiseman, N. J. Phys. 9, 165 (2007).

\bibitem{bohm} D. Bohm and B.J. Hiley \emph{The Undivided Universe} (London:
Routledge,1993). P. R. Holland \emph{The quantum theory of motion}
(Cambridge, Cambridge University Press, 1993).

\bibitem{mismatch} A. Matzkin Found Phys 39 903 (2009).

\bibitem{wavepackets} J. J. Mestayer et al, Phys. Rev. Lett. 99 183003  (2007); A. Buchleitner et al, Phys. Rep. 368 409 (2002).


\bibitem {louisell}See Sec. 1.16 of W. H. Louisell \emph{Statistical properties of
radiation} (Wiley, New York, 1973).


\bibitem {chap}D. Lauvergnat, S. Blasco, X. Chapuisat and A. Nauts, J.
Chem. Phys 126, 204103.



\bibitem {shps}A. Matzkin and V. Nurock, Studies in Hist Phil. Science B 39, 17, (2008).

\end{thebibliography}
\end{document}